\documentstyle[12pt]{article}
\newcommand{\bcn}{\begin{center}}
\newcommand{\beq}{\begin{equation}}
\newcommand{\beqn}{\begin{eqnarray}}
\newcommand{\ecn}{\end{center}}
\newcommand{\eeq}{\end{equation}}
\newcommand{\eeqn}{\end{eqnarray}}

\def\lapprox{\mathrel{\mathop{\kern 0pt <}\limits_{\displaystyle\sim}}}
\def\gapprox{\mathrel{\mathop{\kern 0pt >}\limits_{\displaystyle\sim}}}
\begin{document}
\newpage

\mbox{}
\vspace*{\fill}
\begin{center}
{\LARGE\bf Probing the polarized gluon content } \\

\vspace{2mm}
{\LARGE\bf  of the proton through $ \chi_2$ hadroproduction\footnote{\rm to
be published in the
 proceedings of the  2nd Meeting on Possible
Measurements of Singly Polarized pp and pn Collisions, Zeuthen (1995)}
}\\

\vspace{2em}
\large
 Bernard PIRE
\\
\vspace{2em}
{\it   Centre de Physique Th\'eorique{\footnote {Unit\'e propre 14 du Centre
National
de la Recherche Scientifique.}},}
 \\
{\it Ecole Polytechnique, F91128 Palaiseau,
France}\\
%\today
\end{center}
\vspace*{\fill}
\begin{abstract}
\noindent

Determining how much spin is carried by gluon in a polarized proton is a
fundamental problem which cannot be resolved by completely inclusive deep
inelastic measurements. Hadroproduction of heavy flavors is very sensitive
to the
gluon content of hadrons. We show that $\chi_2 (3555)$ production in
polarized proton-proton collisions is a good candidate reaction to address
this challenging question.

\end{abstract}
\vspace*{\fill}
\newpage

Charmonium spectrum and decays are quite well understood in the framework of
perturbative QCD because these states  are small nonrelativistic systems of
heavy quarks. The production of charmonia in hadron-hadron collisions is
expected
to be dominated by gluon fusion mechanisms. Whereas $ J/\psi$ production
seems to
be subject to important higher twist contributions~\cite{VHBT}, namely
those where
two gluons  from a single hadron interact with a gluon from the other
hadron, the
production~\cite{exp} of $\chi_j $ states is phenomenologically~\cite{VHBT}
 well under control in the simple leading twist picture where two gluons (one
 from each hadron) fuse to give at leading order a $1^+$ $c\bar c$ state with
 zero transverse momentum. Among these states, $\chi_2 (3555)$ is remarkably
sensitive to incident gluon polarizations~\cite{CP} and thus deserves
careful investigation.

Neglecting $O(\alpha_S)$ corrections and internal transverse momenta in the
proton,
the production cross section (at zero transverse momentum) is given by\cite{BR}
\beq
\sigma (x_F \ge 0)  = {{M^2}\over {s}}\int {{dx_1} \over {x_1}}~ G
(x_1,M^2)~ G({{M^2}\over {x_1s}},M^2)
    \sigma_0 (gg \rightarrow \chi_2)
\eeq
where M is the $\chi_2$ mass and
\beq
\sigma_0 (gg \rightarrow \chi_2) = 16 \pi^2 \alpha_s^2 \mid R_P'(0)\mid ^2
/ M^7,
\eeq
 reasonable estimates of the parameters being~\cite{VHBT}, $\alpha_S$ = 0.26
and
$\mid R_P'(0)\mid  / M $ = $ 0.006 Gev^3 $.

The helicity amplitudes for producing the $\chi_2$ meson with $J_z = 0, \pm 1 $
vanish (unless non leading internal transverse momentum effects are kept).
Thus,
$\chi_2$ is only produced with  $J_z =  \pm 2 $ and the sign of $J_z$ is
directly
reminiscent of the helicities ($\mu$ and $\mu'$) of the incident gluons.
Indeed the only non-vanishing
amplitudes are those for
\beq
g(\mu=1) + g(\mu'=-1) \rightarrow \chi_2 (J_z = +2)
\eeq
and
\beq
g(\mu=-1) + g(\mu'=1) \rightarrow \chi_2 (J_z = -2)
\eeq
\vskip.7cm
This makes $\chi_2$ a beautiful probe of the gluon spin content of the proton.
\vskip1cm
\noindent
When both beam and target are polarized, one measures cross sections
$d\sigma_{ij}$ for incident hadron helicities $i$ and $j$, and defines an
 asymmetry as
\beq
A(x_F) = {{{{d\sigma_{++}} \over {dx_F}} - {{d\sigma_{+-}} \over {dx_F}}} \over
{{{d\sigma_{++}} \over {dx_F}} + {{d\sigma_{+-}} \over {dx_F}}}  }
\eeq
which measures how much  gluons remember the spin state of their parents
through
\beq
A(x_F) = {{\Delta G(x_1,M^2)} \over {G(x_1,M^2)}} ~{{\Delta G(x_2,M^2)}
\over {G(x_2,M^2)}}
\eeq
with $x_1$ and $x_2$ detemined from measurable quantities through the usual
relations: $ x_F = x_1 - x_2 $ and $M^2/s = x_1 x_2 $.
\vskip1cm
\noindent
When only the target is polarized, one needs to measure a transmitted
asymmetry,
{\it i.e.} to recognize a $J_z = +2$ $\chi_2$ state from a  $J_z = -2$ one.
This
requires analizing   $\chi_2$ decay channels. The most interesting channel
is the
electromagnetic decay
\beq
\chi_2 \rightarrow J/\psi + \gamma
\eeq
the rate of which is around 13 per cent. In the heavy quark limit, this
transition
is of the electric dipole type, {\it i.e.} it preserves the quark spins.
The angular
decay distribution is thus known as
\beq
{{d\Gamma} \over {d\theta}} = {{3} \over {16 \pi}} (1 + cos^2\theta)
\eeq
and the polarization of the photon along the beam direction is simply given
by~\cite{CP}
\beq
\cal{P_\gamma} = - {{\Delta G(x_2,M^2)} \over {G(x_2,M^2)}}~~.
\eeq

Measuring this outgoing photon polarization is an experimental challenge but
the
rewards are high; it is worth considering it in more details.
\vskip2cm
\noindent
{\it Acknowledgements}.
\vskip.7cm
\noindent
It is a pleasure to thank the organizers and particularly Wolf Dieter Nowak for
 this lively and fruitful workshop.

\end{document}